\documentclass[conference]{IEEEtran}

\IEEEoverridecommandlockouts
\usepackage{cite}
\usepackage{amsmath,amssymb,amsfonts}
\usepackage{algorithmic}
\usepackage{graphicx}
\usepackage{textcomp}
\usepackage{xcolor}
\usepackage{balance}
\usepackage{url}
\usepackage{makecell}

\usepackage{soul}

\def\BibTeX{{\rm B\kern-.05em{\sc i\kern-.025em b}\kern-.08em
    T\kern-.1667em\lower.7ex\hbox{E}\kern-.125emX}}
\begin{document}

\title{Named Entity Recognition for Audio De-Identification\\
}

\author{\IEEEauthorblockN{Guillaume Baril, Patrick Cardinal, Alessandro Lameiras Koerich}
\IEEEauthorblockA{\textit{Department of Software and IT Engineering} \\
\textit{École de Technologie Supérieure, Université du Québec}\\
Montréal, QC, Canada \\
guillaume.baril.1@ens.etsmtl.ca, patrick.cardinal@etsmtl.ca, alessandro.koerich@etsmtl.ca}
}

\maketitle

\begin{abstract}
Data anonymization is often a task carried out by humans. Automating it would reduce the cost and time required to complete this task. This paper presents a pipeline to automate the anonymization of audio data in French. We propose a pipeline, which takes audio files with their transcriptions and removes the named entities (NEs) present in the audio. Our pipeline is made up of a forced aligner, which aligns words in an audio transcript with speech and a model that performs named entity recognition (NER). Then, the audio segments that correspond to NEs are substituted with silence to anonymize audio. We compared forced aligners and NER models to find the best ones for our scenario. We evaluated our pipeline on a small hand-annotated dataset, achieving an F1 score of 0.769. This result shows that automating this task is feasible.
\end{abstract}


\section{Introduction}
The capture of data from customers has been proliferating among many businesses. Companies collect customer data for multiple purposes: improving their recommendations, understanding customer needs, driving decision-making, and much more. An example of data collected by companies in their call centers is the conversations between customers and their employees. Companies mainly collect four categories of data: personal, engagement, behavioral, and attitudinal data. In the particular case of call centers, when customers call, they have an account, and several personal information such as names, social insurance numbers, account numbers, and addresses are required to identify the customer.

In the last few years, there have been many data breaches. "Black hat" hackers steal customers' personal information for many reasons like selling it or committing fraud. There are three main ways for companies to protect themselves from data breaches: limit access to the data, improve employees' security awareness, and patch system vulnerabilities. Another way for companies to protect themselves from data breaches is to delete personal information when it is not needed. For example, in most cases, call center recordings are used for training purposes or to keep track of the employees' behavior. In these cases, the customer's personal information is not essential, and it could be redacted. Also, removing sensitive information from data would allow companies to make the datasets available to the public. Therefore, data de-identification plays an essential role in protecting persons' privacy and also making such data available for researchers. 

Automatic de-identification of data is not a new task. There is a lot of research in this area. One example is de-identifying medical notes \cite{medicalred1, medicalred3, medicalred2} to protect the confidentiality of patients. Another example is speaker anonymization \cite{speaker1, speaker2, speaker3} to hide the speaker's identity. However, there is not a lot of research in audio de-identification. To the best of our knowledge, Cohn {\it et al.}~\cite{cohn_may19} are the only researchers who explored this topic using a pipeline with automatic speech recognition (ASR) and named entity recognition (NER). 

Audio de-identification consists of finding named entities (NEs) in speech and removing them afterward. There are not several end-to-end NER from speech models \cite{ghannay_may18, yadav_may20}, but NER on text is a popular research area \cite{huang_bilstm_crf_word, lample_bilstm_crf_word_char, bert, roberta}. However, these methods are language-dependent, and retraining a model for languages other than English is another research of its own \cite{camembert, flaubert}. 

This paper presents a novel approach for anonymizing audio automatically by removing NEs. We consider a scenario of limited resources both in terms of data and computational resources, which is generally the case for languages other than English. Therefore, our approach does not consider ASR. Instead, we assume that an ASR model has already transcribed audio, and we apply a forced alignment algorithm (FAA) to align each word with the audio recording. Finally, we find the NEs in the transcription and remove them from the audio.
The proposed pipeline to anonymize French audio recordings automatically does not require plenty of data to achieve good performance. The main contributions of this paper are: (i) an annotated French speech corpus with word boundaries; (ii) evaluation of FAAs suitable for the French language; (iii) a NER model trained on a French corpus of financial news; (iv) a pipeline to anonymize French speech automatically\footnote{The link to GitHub will be provided in the final version.
}; (v) a new evaluation metric to assess the performance of audio NER models.

This paper is organized as follows. Section~\ref{sec:prev} presents related works and state-of-the-art FA algorithms and NER architectures. Section~\ref{sec:proposed} describes the proposed pipeline for audio anonymization and FA and NER algorithms employed in the proposed pipeline. The corpora and the experiments carried out on the pipeline created to anonymize audio recordings are presented in Section~\ref{sec:exp}. Finally, the last section presents our conclusions and recommendations for future work.

\section{Previous Works}
\label{sec:prev}
There are few approaches for the audio de-identification task, and they are based on either end-to-end or pipeline models. Pipeline models for audio de-identification usually have three components: an ASR to transcribe audio to text, a text NER that recognizes NEs in the audio transcript, and an alignment component that removes the entities from audio. End-to-end models are based on spoken language understanding models that are trained on audio recordings with NER annotations. Therefore they bypass audio-to-text transcription and de-identify audio recordings directly. In this section, we present some approaches that have been proposed for NER from speech, as well as some recent end-to-end approaches for text NER, as they are part of the pipeline proposed in this paper.

Ghannay {\it et al.}~\cite{ghannay_may18} proposed an end-to-end NER from speech based on SLU models. They used an architecture similar to DeepSpeech 2 \cite{deepspeech2}. The character sequence outputted by the model is composed of the alphabet and nine NE tags. They proposed two strategies to compensate for the lack of data during training. The first strategy consisted of using a multi-task learning approach. Firstly, they trained the model only on the ASR task, and after the training, they reinitialized the softmax layer to consider the nine NE tags. Then, they retrained the model using audio recordings with NER annotations. The second strategy increased the amount of data by annotating audio recordings without NER annotations with a text NER system. The new annotated audio was used in the training phase of the end-to-end model. They trained and evaluated their models on the DeepSUN corpus, which combines four French corpora composed of audio recordings from radio and television emissions. The end-to-end model was compared with a pipeline model for detecting and extracting NEs within a sentence. The end-to-end model outperformed the pipeline on the detection task showing precision, recall and F1 score of 0.76, 0.63 and 0.69, respectively. However, the pipeline model outperformed the end-to-end model on the extraction task, achieving precision, recall and F1 score of 0.57, 0.45 and 0.50, respectively, against 0.49, 0.41 and 0.47 achieved by the end-to-end model. Therefore, the end-to-end model was better to determine if there is a NE in a sentence, but it could not say which words are part of that entity.

Cohn {\it et al.}~\cite{cohn_may19} proposed a pipeline to de-identify recorded conversations between patients and doctors. The pipeline first produces transcripts from the audio using ASR, proceeds by running text-based NER tagging, and then redacts personal health identifiers tokens, using the aligned token boundaries determined by ASR. The NER relies on state-of-the-art techniques for solving the audio NER problem of recognizing entities in audio transcripts. In addition, they leveraged the ASR and use its component of alignment back to audio. They also defined and published a benchmark dataset consisting of a large labeled subset of the Switchboard and Fisher conversational English audio datasets, called SWFI. The best model achieved recall, precision, and F1 score of 0.88, 0.92, and 0.90, respectively, for the audio de-identification task, on the SWFI test set using six NEs.


Yadav {\it et al.}~\cite{yadav_may20} proposed an end-to-end audio NER approach for English speech, which jointly optimizes the ASR and NER tagger components. They also introduced a publicly available NER annotated dataset for English speech, which results from the combination of four speech datasets. Their approach recognizes only three NEs: organization, person, and location. The end-to-end approach is also based on DeepSpeech 2 with modifications in the last layer to adapt it to the NER task. The end-to-end approach achieved precision, recall and F1 score of 0.96, 0.85, and 0.90, respectively, on the test set of their DATA2 dataset. They compared their end-to-end approach with a NER pipeline, which achieved precision, recall and F1 score of 0.83, 0.77, and 0.80, respectively. They stated that they reached a better performance than Ghannay {\it et al.}~\cite{ghannay_may18} because the word error rate (WER) of their model was 2.72\% compared to the 19.96\% of Ghannay {\it et al.}~\cite{ghannay_may18}. Therefore, improving the ASR WER improves the overall performance of the end-to-end model.

Several text NER architectures have been presented in the last years~\cite{huang_bilstm_crf_word,lample_bilstm_crf_word_char,survey_lstm,chiu_bilstm_cnn,bert}. We split them in two categories: LSTM-based and transformer-based. LSTM is one of the best models for natural language processing (NLP) tasks such as text NER because it can learn long-term dependencies. Huang {\it et al.}~\cite{huang_bilstm_crf_word} evaluated simple architectures such as a forward LSTM, a BiLSTM and a CRF and they also proposed a more complex network by combining a BiLSTM with a CRF layer on top. The BiLSTM allows the model to use past and future input features and the CRF layer allows the model to use sentence level tag information. The CRF uses the BiLSTM left-right context representations as input. Another architecture introduced by Lample {\it et al.}~\cite{lample_bilstm_crf_word_char} is a BiLSTM-CRF network combining both word and character-level context. Unlike previous architectures, which only use word-level context, this new architecture have proved to have good performance with little domain specific knowledge \cite{survey_lstm}.
Instead of using a BiLSTM to model character-level information, Chiu and Nichols~\cite{chiu_bilstm_cnn} used a simple CNN architecture with one convolution layer followed by a max pooling. Then, they concatenated the CNN outputs and the word embeddings before feeding them to the BiLSTM. The resulting model is a BiLSTM-CNN. That being said, the best model without any additional data is the word-level BiLSTM-CRF by Huang {\it et al.}~\cite{huang_bilstm_crf_word} with a F1 score of 0.842 and the best model using additional data is the word+character-level BiLSTM-CNN by Chiu and Nichols~\cite{chiu_bilstm_cnn} with a F1 score of 0.916. We believe that the word+character-level BiLSTM-CRF model performed worst than the word-level BiLSTM-CRF model because it has many more parameters and that leads to overfitting.

The bidirectional encoder representation from transformers (BERT)~\cite{bert} is one of the most popular transformer-based language representation model. BERT\footnote{BERT: 13GB of training, NSP \& MLM pre-training tasks, 100M parameters, WordPiece 30k tokenizer, static masking strategy.} uses an encoder-decoder architecture and a baseline and a large architecture have been proposed. These architectures have 12 and 24 layers, 768 and 1024 hidden layers of size 4, and 12 and 16 self-attention heads, respectively. Both models use wordpiece embeddings with a vocabulary of 30 000 tokens as input and output representations and they are trained in two steps. BERT is pre-trained on a large data using two unsupervised tasks: (i) masked LM (MLM) consisting of masking some wordpiece tokens at random and predicting these masked words to understand context in a sentence; (ii) next sentence prediction (NSP) consisting of feeding two sentences to the model and predicting if sentence $A$ follows sentence $B$ or not to understand relation between multiple sentences, which results in a language model. Then, the second step, which does not require as much resources as the first step, is to fine-tune BERT on specific tasks with labeled data. The authors fine-tuned BERT on a text NER dataset~\cite{conll03} of news stories with four NEs: organizations, persons, locations and miscellaneous. The baseline and the large BERT architectures achieved an F1 score of 0.924 and 0.928 on the test set of CoNLL-2003 dataset~\cite{conll03}. 

Liu {\it et al.}~\cite{roberta} have made significant improvements to BERT given origin to RoBERTa\footnote{RoBERTa: 160GB of training, MLM pre-training task, 125M parameters, BPE 50k tokenizer, dynamic masking strategy.}, which stands for a robustly optimized BERT pre-training approach. They improved the masked LM task by using dynamic masking, which changes the masked tokens of each sentence in every training epoch. Also, they found that removing the NSP task matches or improves performance. Then, they used a byte-pair encoding (BPE) tokenizer \cite{bpe} with a 50 000-token vocabulary size. Additionally, they found that BERT was undertrained during the pre-training step. Therefore, they trained their model longer with a bigger batch size. Finally, they used longer input sequences, so the model can learn long-range dependencies. RoBERTa was not evaluated on text NER tasks, but its architecture was used on two French language models~\cite{flaubert,camembert}. FlauBERT~\cite{flaubert} has the same architecture as RoBERTa but it was trained on a French corpus. They collected their data from three main sources: one corpus from the WMT19 shared task \cite{wmt19}, one corpus from the OPUS collection \cite{opus} and three datasets from the Wikimedia projects. After preprocessing, the training corpus was 71 GB in size. CamemBERT~\cite{camembert} is another French language model based on RoBERTa, but there is three main differences: (i) it uses sentence piece tokenization \cite{sentencepiece} with a vocabulary size of 32,000 tokens, which is an extension of BPE and wordpiece and does not require pre-tokenization; (ii) it uses whole-word masking during the MLM task, where the whole original word is masked instead of sub-tokens created by the tokenizer; (iii) it was trained with the French part of the OSCAR corpus~\cite{oscar} consisting of 138 GB of raw text. FlauBERT and CamemBERT achieved similar text classification accuracy of 93.22\% and 93.38\%, respectively~\cite{flaubert}. It is important to note that FlauBERT and CamemBERT were not evaluated on text NER tasks in the original papers.

\section{Proposed Approach}
\label{sec:proposed}
Despite the advantages of end-to-end models, the proposed approach for French speech de-identification is based on a pipeline model due to two practical issues: (i) there is no publicly available pre-trained end-to-end NER model for speech in French, and training SLU models from scratch would require a large amount of data and computational resources; (ii) current end-to-end NER models, in general, do not outperform pipeline models.

\begin{figure*}[ht]
	\centering
		\includegraphics[width=0.7\textwidth]{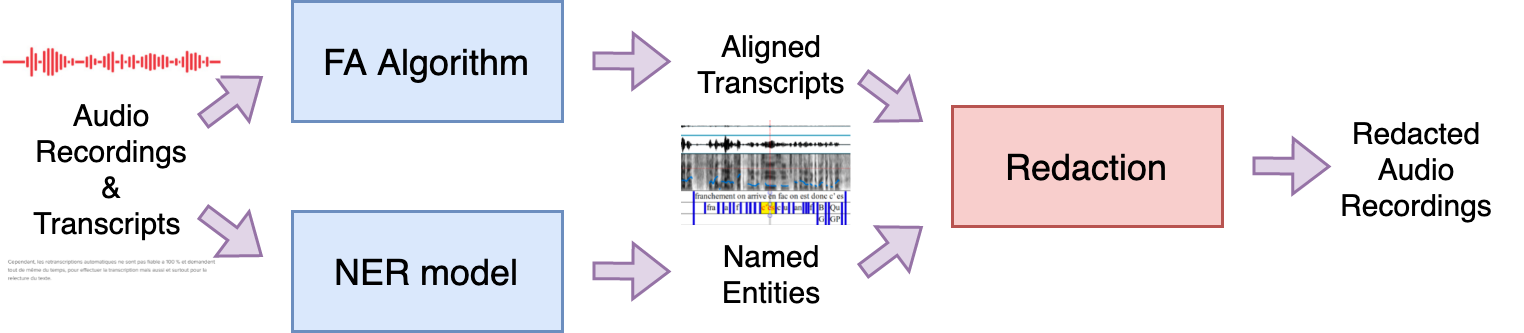} \\
		\caption[Pipeline architecture]{Overview of the architecture of the audio de-identification pipeline\label{fig:pipeline_architecture_overview}}
\end{figure*}

The proposed approach to de-identify French audio recordings consists of a pipeline with three components, as shown in Fig.~\ref{fig:pipeline_architecture_overview}. The first component is a FA algorithm that aligns audio transcriptions  with the corresponding audio. The second component is a text NER model that finds NEs in the audio transcription. Finally, the redaction component substitutes the audio inside the boundaries of the NEs with silence, outputting a redacted audio recording.

The final application aims to de-identifying call center recordings of financial institutions. Therefore, the NEs are related to currencies, locations, money amounts, organizations, and persons.

\subsection{Forced Alignment (FA) Algorithms}
\label{fa_algo_lit_review}
There are several algorithms to align each word of a transcribed audio with the corresponding audio signal. Most of them are based on the hidden Markov model toolkit (HTK) \cite{htkbook}, or Kaldi \cite{kaldi}. Some of them, such as FAVE \cite{fave} are language-dependent and only work in the English language. However, our main interest is in FA algorithms that support the French language, such as the Montreal forced aligner (MFA) \cite{McAuliffe2017}, and the speech phonetization alignment and syllabification (SPPAS) \cite{bigi2015}.

MFA is based on Kaldi, which uses an HMM/GMM architecture with three training phases for acoustic modeling. The first phase consists of training monophone GMMs to generate basic alignment. Then, the second phase uses these models to train triphone GMMs, which consider the context of a phoneme. Also, it uses a decision tree to cluster triphones to avoid sparsity. Finally, based on the triphone models, it uses constrained maximum likelihood linear regression (CMLLR) to calculate speaker-adapted GMMs \cite{CMLLR, fmllr}. 

SPPAS~\cite{bigi2015} is another forced alignment algorithm, and it is based on HTK acoustic models (AM) and language models (LM) and on Julius decoder~\cite{julius}. Training is done in three steps. Firstly, it phonetizes the text with a language-independent algorithm \cite{bigi2016}. This phonetic representation will be used in the following steps. Then, it creates a flat start monophone model. In other words, it will initialize the model by calculating the average of all the features in the corpus and then train the model with the Viterbi algorithm. Finally, it creates a tied-state triphone model, which is a decision tree of triphone HMMs to avoid sparsity. The authors also worked on a Quebec French model \cite{enfrancaissvp}. \cite{bigi2018} evaluated SPPAS on various spontaneous speech corpus in French.

We have evaluated MFA with its pre-trained acoustic model \textit{french\_prosodylab} and its associated dictionary using a beam size of 100 and 8 kHz sample rate, and SPPAS~\cite{bigi2015} with its pre-trained acoustic model named \textit{fra} and its dictionary, using a beam size of 1,000 and a sampling rate of 16 kHz. We did not use Cepstral mean and variance normalization because it can degrade the model performance when the utterances are short, which is our case. The main difference between MFA and SPPAS is that MFA can deal with audio sampled as low as 8 kHz, while SPPAS cannot do that by default because it has been trained on audio sampled at 16 kHz.

\subsection{Text Named Entity Recognition (NER) Models}
We have evaluated two different LSTM-based models\footnote{For both LSTM-based models, we use the code created by G. Lample, available at \url{https://github.com/glample/tagger} with character layer size of 25, word layer size 100, dropout 0.5, SGD learning rate of 5$\times$10$^{-3}$, and the number of epochs between 15 and 30.} and a pre-trained transformer model for text NER. The LSTM-based models were fully trained on the FrenNER dataset using the stochastic gradient descend algorithm. It is important to notice that there are not many resources for French compared to English, which restrain us from using pre-trained embeddings or gazetteers. The first LSTM-based model is the word-level BiLSTM-CRF \cite{huang_bilstm_crf_word}, which achieved the highest F1 score on the CoNLL-2003 dataset. The second LSTM-based model is the word+character-level BiLSTM-CRF \cite{lample_bilstm_crf_word_char}, which has shown good performance with little domain-specific knowledge. This model achieved the second-best F1 score on the CoNLL-2003 dataset, with and without additional training data.
We have used the same layer size for both BiLSTMs because it does not significantly affect model performance \cite{huang_bilstm_crf_word}.

Finally, we have evaluated the CamemBERT transformer model \cite{camembert}\footnote{We have used the base model available with Huggingface with almost all the default hyperparameters except for the batch size of 16, the learning rate of 5$\times$10$^{-5}$, and 5 training epochs.}. We have chosen CamemBERT over FlauBERT because it performs better in almost all tasks. Besides, CamemBERT is pre-trained on more data than FlauBERT, which should enhance its performance on downstream tasks such as text NER. We used the pre-trained encoder and we fine-tuned it on the text NER task using the FrenNER dataset.  

\section{Experimental Results}
\label{sec:exp}
The proposed pipeline and its components were evaluated on two French corpora related to the NER task. The Nijmegen Corpus of Casual French (NCCFr) was used to train and evaluate the FA models and the final pipeline for audio de-identification. The French corpus for NER and Relation Extraction of financial news (FrenNER) is a text corpus used to train and assess the text NER models. 

\begin{figure}[ht]
	\centering
		\includegraphics[width=0.49\textwidth]{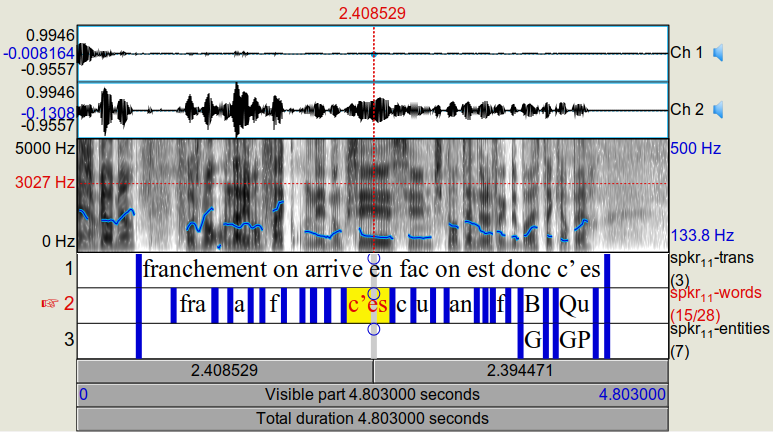}
		\caption{Example of one phrase annotated word by word with Praat\label{fig:annotation_praat}}
\end{figure}

NCCFr~\cite{TORREIRA2010201} is a speech corpus, which contains 36 hours of transcribed conversations of 46 different speakers from multiple regions of France. This corpus was manually annotated to determine word boundaries. For such an aim, we have split each corpus interval corresponding to a phrase of a single speaker. Then, we have randomly chosen phrases potentially containing NEs, and we annotated 381 seconds\footnote{The manual annotation was carried out using Praat\cite{Boersma2020}.}. Fig.~\ref{fig:annotation_praat} shows an example with annotated words' boundaries and each NE.

FrenNER~\cite{jabbari-etal-2020-french} is a text corpus made of 130 manually annotated news articles from 40 daily French newspapers. The corpus contains 6,736 entities of 26 different types. We have preprocessed the dataset to: (i) replace some characters such as \textit{\textbackslash t} or \textit{\textbackslash u2009} with spaces; (ii) remove multiple spaces and leading determiner from NE annotations; (iii) modify or delete some entity types, which are either irrelevant or low frequent. We only kept currencies, locations, money amounts, organizations and persons, and we modified these entities to fit into one of the five types: (i) \textit{shareholderships} and \textit{financing} have money amounts and currencies. So, we decided to change some of them into money amounts and currencies, and we deleted the others because they were irrelevant (\textit{d'actionnaires, entre au capital, lever, investis, etc.}); (ii) \textit{geopolitical entities} were converted into locations and organizations, or they were deleted because they were irrelevant (\textit{les gouvernants, etat, g20, etc.}); (iii) \textit{world regions}, \textit{countries}, \textit{local regions}, and \textit{cities} were all converted into locations because it was not necessary to split them into sub-types; (iv) \textit{agents}, \textit{associations}, \textit{medias}, and \textit{companies} were all converted into organizations also because it was not necessary to split them into sub-types. Finally, we split the news articles into 4,424 sentences. These sentences were randomly split into ten groups and used in a 10-fold cross-validation scheme.

\subsection{Evaluation Metrics}

The evaluation of a FA model is done by calculating the accuracy on a corpus $C$ defined as:

\begin{equation}\label{eq:eval_fa}
    \text{Accuracy}  = \frac{1}{|C|}\sum_{p, g \in C}\sum_{p_i, g_i \in p, g}\delta_t(p_i, g_i)
\end{equation}

\noindent where each pair $p,g$ are the model prediction and the gold standard, respectively, $p_i$ and $g_i$ represents the $i^{\text{th}}$ word boundary and $\delta_t$ is a function returning 1 if $p_i$ is correctly aligned given a tolerance $t$. Eqs.~\eqref{eq:delta_s} and~\eqref{eq:delta_o} show the two $\delta_t$ functions that can be used in Eq.~\eqref{eq:eval_fa}. $\delta_t^s$ 
is the absolute difference between the prediction and the gold standard (\textit{std}), while $\delta_t^o$ has the tolerance $t$ applied only inside the boundaries (\textit{outer}).

\begin{equation}\label{eq:delta_s}
    \delta_t^s(p_i, g_i) = 
        \begin{cases}
         1 &\quad\text{if } |p_i^0-g_i^0| \leq t \text{ \& } |p_i^1-g_i^1| \leq t \\
         0 &\quad\text{otherwise}\\
        \end{cases}
\end{equation}
\begin{equation}\label{eq:delta_o}
    \delta_t^o(p_i, g_i) = 
        \begin{cases}
         1 &\quad\text{if } p_i^0 \leq g_i^0 + t \text{ \& } g_i^1 - t \leq p_i^1  \\
         0 &\quad\text{otherwise}\\
        \end{cases}
\end{equation}

\noindent where $p_i^0$ and $p_i^1$ are the time stamps when the $i^{\text{th}}$ word starts and ends, respectively. The same  notation is used for $g_i$. 

Eqs.~\eqref{eq:precision} to~\eqref{eq:f1score} show the three metrics for evaluating NER models:

\begin{equation}\label{eq:precision}
    \text{Precision} = \frac{\textit{TP}}{\textit{TP} + \textit{FP}}
\end{equation}

\begin{equation}\label{eq:recall}
    \text{Recall} = \frac{\textit{TP}}{\textit{TP} + \textit{FN}}
\end{equation}

\begin{equation}\label{eq:f1score}
    \text{F1 Score} = \frac{2*\text{Precision}*\text{Recall}}{\text{Precision}+\text{Recall}}
\end{equation}

\noindent where \textit{TP} is when the model predict the right type of a NE, \textit{FP} is when the model predict a NE when there is none or does not predict the right type of a NE, \textit{FN} is when the model does not find a NE.

The evaluation of the final pipeline uses a new metric called No Type Error (NTE) that computes precision, recall and F1 score where misclassified NE types does not count as \textit{FP} or \textit{FN} if any type of NE is found. Therefore, \textit{TP}, \textit{FP} and \textit{FN} are computed using the \textit{outer} function $\delta_t^o$ where \textit{t} is the tolerance as: 

\begin{equation}
    \textit{TP} = \sum_{p, g \in K}\sum_{p_i, g_i \in p, g}\delta_t^o(p_i, g_i)
    \label{NTE:TP}
\end{equation}
\begin{equation}
    \textit{FN} = \sum_{p, g \in K}\sum_{p_i, g_i \in p, g}1-\delta_t^o(p_i, g_i)
    \label{NTE:FN}
\end{equation}
\begin{equation} 
    \textit{FP} =  \sum_{p, g \in K}\sum_{p_i, g_i \in p, g} \max(0, \min(1, g_i^0 + g_i^1))
    \label{NTE:FP}
\end{equation}

\noindent where \textit{K} is the set of predictions made by the model with its corresponding gold standard. \textit{TP} is equal to the number of NEs found and correctly aligned. \textit{FN} is equal to the number of NEs found and not correctly aligned plus the number of gold standard annotations without any corresponding prediction. Since we represent the absence of prediction for a certain gold standard as $p_i^0 = 0$ and $p_i^1 = 0$, the \textit{outer} function will always return 0 in that case. Finally, \textit{FP} is equal to the number of predictions without any corresponding gold standard annotation. Similarly, the absence of a gold standard for a certain prediction is represented as $g_i^0 = 0$ and $g_i^1 = 0$.

\subsection{Evaluation of FA algorithms}
Table~\ref{tbl:tolerance_fa} shows the accuracy of 
aligning the corpus with both MFA and SPPAS algorithms for multiple tolerances.
For SPPAS, there is a difference in accuracy of less than 1\% between the \textit{std} and the \textit{outer} functions starting with a tolerance of 0.60 seconds. For MFA, it starts with a tolerance of 0.20 seconds. Also, with the \textit{outer} function, SPPAS has less than 1\% increase in accuracy over the previous tolerance starting at 0.50 seconds. For MFA, it is even better at 0.30 seconds.

\begin{table}[!ht]
\centering
\caption{Accuracy of FA models at different tolerances}
\label{tbl:tolerance_fa}
\begin{tabular}{|c|c|c|c|c|}
    \hline
        {\bf Tolerance} & \multicolumn{2}{c|}{\bf SPPAS} & \multicolumn{2}{c|}{\bf MFA} \\
        \multicolumn{1}{|c|}{\bf (s)} & \multicolumn{1}{c}{\bf [std]} & \multicolumn{1}{c|}{\bf [outer]} & \multicolumn{1}{c}{\bf [std]} & {\bf [outer]} \\
    \hline
        $\leq$0.01 & 0.045 & 0.360 & 0.060 & 0.298 \\
    \hline
        $\leq$0.10 & 0.745 & 0.852 & 0.889 & 0.911 \\
    \hline
        $\leq$0.20 & 0.846 & 0.899 & {0.950} & 0.958 \\
    \hline
        $\leq$0.25 & 0.873 & 0.917 & 0.966 & 0.969 \\
    \hline
        $\leq$0.30 & 0.892 & 0.928 & 0.975 & {0.977} \\
    \hline
        $\leq$0.40 & 0.919 & 0.943 & 0.983 & 0.985 \\
    \hline
        $\leq$0.50 & 0.937 & {0.951} & 0.989 & 0.989 \\
    \hline
        $\leq$0.60 & {0.947} & 0.955 & 0.992 & 0.992 \\
    \hline
\end{tabular}
\end{table}

In French, the oral flow is around 200 words per minute~\cite{200_mots_minutes} or approximately three words per second, and the average number of syllables per word is around 1.25~\cite{etudedistributionsyllabes}. Therefore, we have about four syllables per second, which means each syllable takes an average of 0.25 seconds. Also, NEs are typically composed of multiple syllables. Thus, we decided to use a tolerance of 0.25 seconds for the last phase, which is equal to one syllable.
Finally, we choose the MFA algorithm for our pipeline because it has better accuracy than SPPAS.

\subsection{Evaluation of Text NER Models}
The second experiment aims to evaluate the
best NER model for the pipeline. Therefore, the
three NER models are trained and evaluated on FrenNER dataset.
We used 10-fold cross-validation to evaluate the robustness of the models.
The metrics used for training are the loss on the training partition, and the precision, the recall and the F1 score on the validation partition. 
Table \ref{tbl:total_test} shows the average performance of the models on the test partitions of FrenNER using the conventional metric (Eqs.~\eqref{eq:precision}-\eqref{eq:f1score}) as well as the NTE metric where \textit{TP}, \textit{FN}, and \textit{FP} of Eqs.~\eqref{eq:precision} and~\eqref{eq:recall} are computed by Eqs.~\eqref{NTE:TP}-\eqref{NTE:FP}.

\begin{table}[ht]
\centering
\setlength{\tabcolsep}{2.5pt}
\caption{Performance of each model on the FrenNER considering the conventional (1\textsuperscript{st} lines) and the NTE (2\textsuperscript{nd} lines) metrics. Average and standard deviation values. \label{tbl:total_test}}
\begin{tabular}{|l|c|c|c|}
    \hline
        {\bf Model} & {\bf Precision} & {\bf Recall} & {\bf F1 Score} \\
    \hline
        Word BiLSTM-CRF & 0.808 ± 0.029 & 0.768 ± 0.038 & 0.786 ± 0.013 \\
         & 0.853 ± 0.032 & 0.811 ± 0.038 & 0.830 ± 0.012 \\
    \hline
        Word+Char BiLSTM-CRF & 0.800 ± 0.051 & 0.726 ± 0.059 & 0.757 ± 0.017 \\
         & 0.858 ± 0.042 & 0.779 ± 0.073 & 0.813 ± 0.022 \\
    \hline
        CamemBERT & 0.865 ± 0.014 & 0.885 ± 0.024 & 0.874 ± 0.013 \\
         & 0.889 ± 0.016 & 0.910 ± 0.023 & 0.899 ± 0.013 \\
    \hline
\end{tabular}
\end{table}

Surprisingly, the word+char-level BiLSTM-CRF performed worst than the word-level BiLSTM-CRF. Nonetheless, CamemBERT performed better than both BiLSTM-CRF models with a F1 score of 0.899.
Therefore, we decided to integrate CamemBERT in our pipeline.

Since our final goal is to create a pipeline to anonymize call recordings, we consider less severe to remove more information than removing less information. Thus, improving recall while decreasing precision is acceptable. Therefore, we tried to add a confidence threshold over the "not an entity" probability. For example, with a confidence threshold of 0.5, if the "not an entity" label has a probability of less than 0.5, we change it to 0 and apply another softmax to recalculate the probability of each label. Then, we choose the most likely entity type.

We have carried out some experiments for choosing a threshold that affects the NTE F1 score less than 0.01. A threshold of 0.90, yields a NTE F1 score of 0.893, a NTE precision of 0.835, and a NTE recall of 0.959. In other words, such threshold allows finding almost 96\% of all NEs in our dev set, and 84\% of our predictions are good. The average performance of CamemBERT model with 0.9 confidence score on the test set is shown in Table~\ref{tbl:final_perf_camem}. If we compare with Table \ref{tbl:total_test}, the model lost 4.4\% precision to gain 5.0\% recall without affecting the F1 score. So, this solution seemed promising. Also, the difference between the NTE results are almost the same.

\begin{table}
\centering
\caption{CamemBERT performance on FrenNER dataset with a confidence threshold of 0.9 for each NE. Average and standard deviation values.\label{tbl:final_perf_camem}}
\begin{tabular}{|l|c|c|c|}
    \hline
        {\bf Entity Type} & {\bf Precision} & {\bf Recall} & {\bf F1 Score} \\
    \hline
        Person & 0.939 ± 0.020 & 0.963 ± 0.017 & 0.951 ± 0.016 \\
    \hline
        Currency & 0.848 ± 0.048 & 0.949 ± 0.035 & 0.895 ± 0.039 \\
    \hline
        Location & 0.878 ± 0.032 & 0.940 ± 0.014 & 0.908 ± 0.019 \\
    \hline
        Money Amount & 0.818 ± 0.029 & 0.969 ± 0.021 & 0.887 ± 0.019 \\
    \hline
        Organization & 0.759 ± 0.024 & 0.914 ± 0.024 & 0.829 ± 0.017 \\
    \hline
        \multicolumn{1}{|c|}{Total} & 0.821 ± 0.015 & 0.935 ± 0.015 & 0.874 ± 0.010 \\
    \hline
        \multicolumn{1}{|c|}{NTE} & 0.842 ± 0.016 & 0.960 ± 0.012 & 0.897 ± 0.010 \\
    \hline
\end{tabular}
\end{table}

In summary,
we have chosen MFA as our FA algorithm because it has the best accuracy over the two FAAs. We have chosen CamemBERT as our NER model because it was pre-trained on a large dataset, it outperforms other models and is faster to be fine-tuned on new data than training a LSTM-based models from scratch.

\subsection{Anonymazing Audio Recordings}

The final architecture of the proposed pipeline (Fig.~\ref{fig:pipeline_architecture_overview}) integrate the MFA algorithm as FA algorithm and CamemBERT as NER model. The audio recordings and the corresponding transcripts are given to the NER model and FA models, respectively. They output the aligned transcripts and the NEs use by the redaction function. Then, the evaluation function compares the redacted audio recordings with the gold annotations in order to evaluate the performance of the pipeline.
Table~\ref{tbl:pipeline_results} shows the performance of the pipeline on 85 examples of the FrenNER corpus, considering a 
tolerance $t$ of 0.25 seconds and without 
a confidence score.

\begin{table}[ht]
\centering
\caption{Pipeline performance on the NCCFr speech corpus using NTE metric.\label{tbl:pipeline_results}}
\begin{tabular}{|c|c|c|}
    \hline
        \textbf{Precision} & \textbf{Recall} & \textbf{F1 score} \\
    \hline
        0.985 & 0.631 &  0.769 \\
    \hline
\end{tabular}
\end{table}

Unfortunately, even though experiments on the FrenNER dataset were promising, the confidence score did not improve the performance of the pipeline on the NCCFr speech corpus. The confidence threshold over the "not an entity" label has no impact on the performance of our pipeline. It means that the model is always certain that the word it predicted as an entity is one. Combined with the fact that the precision is very high and that the recall is very low on the speech corpus, we suspect the speech corpus to be too small. 

Since the training and validation sets both contains the same NEs, it is hard to say if the models overfit the training set. This assumption has been confirmed when evaluating the pipeline on the speech dataset. Indeed, we obtained results comparable to Ghannay {\it et al.}~\cite{ghannay_may18}. However, we need to keep in mind that their model performs a more complicated task, ASR instead of FA. So, it is normal that their pipeline performance is worse than ours. Like us, their precision is way higher than their recall. While the difference between their precision and recall is around 26\%, our difference is about 56\%.



We believe that the main source of errors is the syntax of the sentences. For instance, the entities were successfully found (in bold) by the pipeline in these two sentences: \textit{"ben la \textbf{Lazio} c' est le hum le club de \textbf{Mussolini}"} and \textit{"et ils montraient justement un un chef de \textbf{Sochaux} et un chef de de \textbf{Lyon}"}. As we can see, the sentence syntax is good even though there are filling words or stuttering. On the other hand, the pipeline could not detect NEs in these two sentences, where the syntax is not good: \textit{"parce qu' après enfin moi moi les autres questions \textbf{Al Gore} son prix \textbf{Nobel} de la paix je m' en fous"}, and \textit{"hum le club le plus raciste quoi de l' \textbf{Italie} un des plus racistes en tous cas"}. They look like two merged sentences. In these situations, the model cannot detect NEs because the NER model was trained on written text and not on speech.

Nevertheless, the performance of the NER model could be improved by an algorithm that splits sentences into phrases when the syntax is not good, so that the input will be more similar to written text. Also, by training the NER model directly on speech with NER annotations, the model will learn examples with that kind of noise. An improved NER model would enhance the pipeline's overall performance since there is no error due to the FA algorithm.


\section{Conclusion}
\label{sec:concl}
This paper presented a novel approach to anonymizing French audio by removing NEs. The proposed pipeline combines a FA algorithm followed by a NER model. Indeed, we did force alignment instead of ASR due to a lack of resources to train an ASR model for the French language. In an environment where data is limited, we recommend to use an already trained model.

In this paper, we began by comparing two FA algorithms on our speech corpus with word-level boundaries in French and chose MFA because it achieved the best accuracy. Next, we trained LSTM-based models using character-level and word-level embeddings and transformers for NER on a French corpus of financial news. Finally, we compared them and chose CamemBERT as our NER model.

The proposed pipeline consists of three steps: (i) aligning the transcription with the audio with the MFA algorithm; (ii) finding NEs in the transcription with a CamemBERT model; (iii) redacting the audio by replacing the NEs with noise. The proposed pipeline for audio NER achieved an F1 score of 0.769 with a precision of 0.985 and a recall of 0.631 on a speech corpus containing annotated NEs.

In future work, we think the first thing to do is to improve the training set by creating a more extensive speech corpus with NER annotations. Indeed, one limitation of this research is the size of our speech corpus. Then, the new speech corpus could be used to train the NER model to be more robust to syntax errors, and it could be trained to detect custom NEs (e.g., bank account, SIN). Another path to explore would be to replace FA with ASR and to train an end-to-end NER model from speech. Given a good training dataset and enough resources, the performance of the end-to-end model should be better than the pipeline with separately trained components.

\balance
\bibliographystyle{IEEEtran}
\bibliography{biblio}

\end{document}